# Political Strategies to Overcome Climate Policy Obstructionism

*By Sugandha Srivastav[1,2,*] and Ryan Rafaty[1,2,3]*

## Abstract

Great socio-economic transitions see the demise of certain industries and the rise of others. The losers of the transition tend to deploy a variety of tactics to obstruct change. We develop a political-economy model of interest group competition and garner evidence of tactics deployed in the global climate movement. From this we deduce a set of strategies for how the climate movement competes against entrenched hydrocarbon interests. Five strategies for overcoming obstructionism emerge: (1) Appeasement, which involves compensating the losers; (2) Co-optation, which seeks to instigate change by working with incumbents; (3) Institutionalism, which involves changes to public institutions to support decarbonization; (4) Antagonism, which creates reputational or litigation costs to inaction; and (5) Countervailance, which makes low-carbon alternatives more competitive. We argue that each strategy addresses the problem of obstructionism through a different lens, reflecting a diversity of actors and theories of change within the climate movement. The choice of which strategy to pursue depends on the institutional context.

† For thoughtful feedback at various stages of this paper's development, we thank Cameron Hepburn, Jacquelyn Pless, William O'Sullivan, Matthew Ives, Sam Fankhauser, Thomas Hale, Joris Bücker, Marion Leroutier, Tim Dobermann, Linus Mattauch, Mike Thompson and INET-Oxford EOS seminar participants. Financial support from the Oxford Martin School Programme on the Post-Carbon Transition is gratefully acknowledged.

1. Smith School of Enterprise and the Environment, University of Oxford

2. Institute for New Economic Thinking at the Oxford Martin School

3. Climate Econometrics, Nuffield College, University of Oxford

*Corresponding author: sugandha.srivastav@ouce.ox.ac.uk

# 1  Introduction

Great socioeconomic transitions involve significant shifts in power. Such was the case for universal suffrage, the abolition of slavery and the end of apartheid. The transition to a post-carbon society will not be different.

Energy systems built around hydrocarbons will have to transition to a zero-carbon paradigm which will entail large shifts in the composition of firms and economic activity. This will inevitably create winners and losers, even if it society as a whole is better off. The existential politics of the post-carbon transition (Colgan, Green and Hale 2020), notably the $10 trillion worth of assets at risk of stranding (Mercure et al. 2018; Tong et al. 2019), makes it particularly prone to obstructionism by entrenched interests.

The climate change countermovement (CCCM) has received growing scholarly attention in recent years (e.g. Brulle 2014; Farrell 2016). The CCCM lobby consists of industry associations, carbon-exposed firms, utilities, workers, unions, corporate-funded think tanks and state-owned enterprises who engage in tactics to weaken climate policies rather than adapt to them. Finding ways to address this obstructionism is important, not only because climate change will affect inequality, conflict, migration, economic development and governance, but also because progress has been stalled in large measure by lobbying and inertia in the political system (Stokes 2020).

The corollary to an active CCCM lobby is the climate movement. The strategic operations of the climate movement have received relatively scant attention in the lobbying literature. To address this gap, we develop a framework that documents five key strategies to overcome obstructionism:

- **Antagonism**, which increases the reputational and economic costs of participating in obstructionism and "business as usual" activities;
- **Appeasement**, which offers monetary relief, retraining and restitution to the losers of the transition;
- **Co-optation**, which seeks change from within by co-opting the opposition to reform and diversify their business model;
- **Institutionalism**, which involves regulatory and structural changes at the level of public institutions to make obstructionism harder; and
- **Countervailance**, which bypasses direct confrontation with political opponents by supporting alternative technologies and strengthening their disruptive market potential.

Each strategy advances a different theory of change, contains distinct tactics and is best suited to different actors (Figure 1). We validate our framework by collecting evidence on the climate movement's activities and categorising that by the five strategies (see database in Supplementary Material). Finally, we develop a political economy model of interest group competition and show how the five strategies, and the tactics within them, change a politician's incentives to enact stronger policy.

We find that the choice of strategy is sensitive to three macro-structural parameters: (i) "democratization", which we define as the bargaining power of citizens relative to corporations (ii) "climate consciousness" which is the bargaining power of citizens who support climate policy relative to those who are against it, and (iii) "green business interests" which is the bargaining power of businesses that support climate policy relative to those that are against it. Once deployed, the strategies themselves affect these variables creating feedback dynamics (Farmer et al. 2019).

Much of the existing literature in climate politics focuses on international climate negotiations. Relatively few studies have investigated how domestic politics and interest group competition constrain climate policy (Keohane 2015). Studies that build on this line on inquiry include Aklin and Urpelainen (2013), Meckling (2019), Brulle (2014, 2019), Farrell (2016), Brulle (2018), Gullberg (2008), McKie (2019), Stokes (2020), and Mildenberger (2020). Our aim is to further build on this literature and make sense of disparate claims on the best path forward towards decarbonisation and overcoming obstructionism.

The rest of this article is structure as follows: Section 2 discusses the issue of climate policy obstructionism and the various forms it takes, Section 3 introduces our theoretical framework which coceptualises a politician's incentives to increase climate ambition and how the five strategies influence this, Section 4 discusses the five strategies in detail with a US case study and Section 5 looks at strategy choice. .

## Figure 1. Five Political Strategies

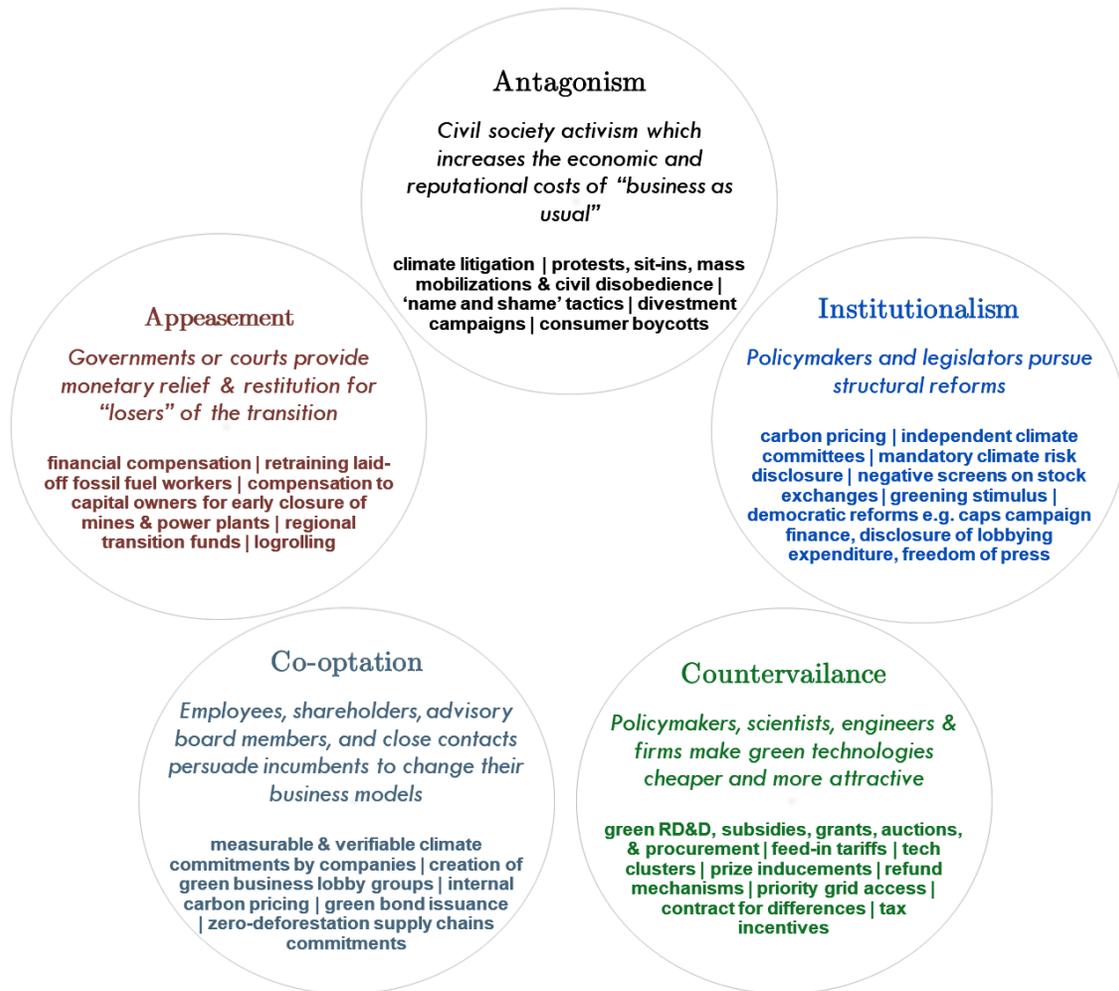

**Antagonism**

*Civil society activism which increases the economic and reputational costs of "business as usual"*

**climate litigation | protests, sit-ins, mass mobilizations & civil disobedience | 'name and shame' tactics | divestment campaigns | consumer boycotts**

**Appeasement**

*Governments or courts provide monetary relief & restitution for "losers" of the transition*

**financial compensation | retraining laid-off fossil fuel workers | compensation to capital owners for early closure of mines & power plants | regional transition funds | logrolling**

**Institutionalism**

*Policymakers and legislators pursue structural reforms*

**carbon pricing | independent climate committees | mandatory climate risk disclosure | negative screens on stock exchanges | greening stimulus | democratic reforms e.g. caps campaign finance, disclosure of lobbying expenditure, freedom of press**

**Co-optation**

*Employees, shareholders, advisory board members, and close contacts persuade incumbents to change their business models*

**measurable & verifiable climate commitments by companies | creation of green business lobby groups | internal carbon pricing | green bond issuance | zero-deforestation supply chains commitments**

**Countervailance**

*Policymakers, scientists, engineers & firms make green technologies cheaper and more attractive*

**green RD&D, subsidies, grants, auctions, & procurement | feed-in tariffs | tech clusters | prize inducements | refund mechanisms | priority grid access | contract for differences | tax incentives**

## 2    Climate Policy Obstructionism

The history of climate policy reveals the extent to which it has been a tug-of-war between different interest groups (Stokes 2020). The global policy landscape is replete with examples of the reversal of climate commitments such as the Australian government's removal of a carbon price only two years after its enactment, the Bolsanaro government's accelerated focus on land-grabbing across the Amazon and Cerrado biomes after years of effectively curbing deforestation (Rochedo et al. 2018) and the US's participation in the Paris Climate Accord which vacillates with which party is in power. The persistent difficulty in phasing out global fossil fuel subsidies is a testament to the degree of hysteresis within the political arena (Skovgaard and van Asselt 2018).

The CCCM lobby in the US has swayed politicians through several different tactics. This includes offering politicians lucrative private sector roles after office (Blanes i Vidal, Draca, and Fons-Rosen 2012), strategically leveraging tax-free corporate philanthropy (Bertrand et al.

2020; Brulle 2018), threatening politicians with competition if they do not acquiesce to demands (Stokes 2020; Dal Bó and Di Tella 2003; Chamon and Kaplan 2013), influencing voters through funding advocacy institutions that promote climate skepticism (DellaVigna, Durante and La Ferarra 2016; Farrell 2016; Farrell 2019), and inserting representatives into regulatory institutions, such as the Environmental Protection Agency to dilute climate policy (Leonard 2019).

CCCM lobbying dwarfs climate movement lobbying on all dimensions including the diversity of tactics, the cultivation of deep political networks (Farrell 2016), and the extent of expenditure (Brulle 2018; Ard, Garcia, and Kelly 2017). For example, lobbying expenditure by the CCCM in the US Congress between 2000-2016 was over USD 2 billion (4% of total lobbying expenditure), which is an order of magnitude higher than the political expenditures of environmental organizations and renewable energy companies (Brulle 2018).

However, the effects of CCCM lobbying extend well beyond the paradigmatic US case. Patterns of obstructionism are manifest in other major fossil-fuel producing countries. For example, in 2013 an estimated one-third of media coverage of climate change in Australia was biased in favour of climate scepticism, with disinformation campaigns openly sponsored by media mogul Rupert Murdoch (Bacon 2013). In India, the government's majority stake in Coal India Limited, the world's largest coal company, creates perverse incentives. In China, provincial politics is tilted in favour of high-carbon prestige projects (Nelder 2021). Even in the European Union, which is considered an innovator in climate policy, carbon-intensive industry associations have actively endorsed the emissions trading scheme (ETS) during periods of reform but have used it as a Trojan Horse to pre-empt stricter regulations. Industry has also negotiated substantial exemptions such as the grandfathering of free allowances and the carbon leakage list which exempts trade-exposed carbon-intensive industries from a carbon price all together (Markard and Rosenbloom 2020).

Passing legislation for decarbonization is difficult because of the sheer value of fossil fuel assets that will be impacted. In monetary terms, the situation is not dissimilar to the abolition of slavery. Slaves made up almost one-fifth of household "assets" back in 1860 and, like fossil fuels, were estimated to be worth around 10-20 trillion USD (Hayes 2014). Abolitionists had to deploy a range of tactics to overcome obstructionism.

Several reasons may explain the CCCM's superior political organization: (i) by virtue of its incumbency, it has greater material resources and political connections at its disposal; (ii) the CCCM lobby is a tightly defined group of actors while the climate movement is relatively more dispersed, making organization costlier; and (iii) existing laws and institutions cater to a high-carbon paradigm which creates inertia in the reform process.

## 3   Political Economy Model

To explore interest group competition, we develop a simple political-economy framework that models how a politician's incentives to enact more stringent climate policy are affected by different agents and institutional factors. While the literature has looked at political competition from the lens of "green" vs. "brown" governments (Aklin and Urpelainen 2013), we extend it to the case of citizens vs. business interest groups (first tier) and, climate conscious citizens and businesses vs. anti-climate citizens and businesses (second tier).

In the model, we assume a politician selects the level of policy ambition, $x$, such that she maximizes the *perceived* welfare, $W$, of citizens and business interest groups (Equation 1). The politician's chance of election or re-election increases in $W$.[1] The politician cares about citizens as they supply votes and businesses since they supply campaign finance.

In our model $x$ represents climate ambition i.e. the target level of emissions reduction. However, in other applications, $x$ may represent the ambition to universalize access to free healthcare, gain autonomy from a subjugating party, or reform the food industry.

$$W(x) = \alpha[\beta_1 W_G^C + (1 - \beta_1)W_F^C] + (1 - \alpha)[\phi W_G^B + (1 - \phi)W_F^B] \qquad (1)$$

$$\underbrace{\qquad\qquad\qquad\qquad}_{\text{Citizenry}} \qquad \underbrace{\qquad\qquad\qquad\qquad}_{\text{Businesses}}$$

$a, \beta_1, \phi \in [0,1]$ describe the relative bargaining power of citizens versus businesses, climate conscious citzens versus anti-climate citizens, and green business interests vs. CCCM business interests respectively.[2] The perceived welfare of $G$ citizens/businesses increases with greater policy ambition ($W_G'(x) > 0$) and decreases for $F$ citizens/businesses ($W_F'(x) < 0$).[3]

Citizen and business interests are considered separately to capture numerous cases of divergent interests. For example, the interests of the youth who are very active in the climate movement have little overlap with that of large business interests.

We focus on *perceived* welfare because the true level of welfare an agent experiences in response to different scenarios may differ from how the agent perceives the matter ex ante, due to misinformation and biases (Druckman and McGrath 2019; Mildenberger and Tingley 2019). In the case of climate change, evidence shows that weather extremes and the promulgation of scientific information do little to change aggregate opinions. Instead, political mobilization by elites and advocacy groups is critical in influencing climate change concern (Brulle, Carmichael and Jenkins 2012).

---

[1] In the case of countries without democratic elections, this can be rephrased as a politician's "ability to retain power".

[2] Consensus democracies such as those of the Nordic countries, or semi-direct representative democracies such as that of Switzerland, will have a relatively high value of $\alpha$. Where there is a strong revolving door between industry and government, such as in the United States, $\alpha$ is lower. In China, where citizens cannot vote but still play a role insofar as they can leverage implicit threats of civil disobedience, $\alpha$ is even lower.

[3] For simplicity we assume there is no neutrality for firms or citizens in relation to how perceived welfare will change in response to climate ambition. This can be modelled but it will not change the core conclusions.

A politician's incentives to increase policy ambition to advance a social movement's agenda can be increased via the five strategies whose tactics change different parts of the politician's objective function. From a static perspective, the choice of strategy is sensitive to initial conditions related to democratization ($\alpha$), climate consciousness ($\beta_1$) and green business incentives ($\phi$). From a dynamic perspective, the strategies start to influence these parameters. Table 1 gives an example of how initial conditions influence strategy choice.

Table 1.  The Sensitivity of Strategies to Initial Conditions

| Initial Conditions (*If*): | Goal (*Then*): | Strategy & Tactic (*By*): |
|---|---|---|
| Citizens are against policy & citizens have at least as much bargaining power as businesses<br>  e.g. deindustrialised mining towns | Increase $\beta_1$ | Financial compensation to coal workers and regional transition funds (*appeasement*) |
| Green business interests are weak and corporations have more bargaining power than citizens.<br>  e.g. US Congress where CCCM interests exert large influence | Increase $\phi$ | Incentivize dirty firms to become clean via:<br>- business model reform and executive incentives (*co-optation*);<br>- tax breaks for clean tech, R&D support, grants (*countervailance*)<br>- financial compensation to capital owners (*appeasement*) |
| Citizens are for policy but have less bargaining power than businesses<br><br>e.g. Germany where a climate conscious citizenry contends with a powerful CCCM lobby | Increase $\alpha$<br><br>Increase $\phi$ | Make it an electoral liability to ignore the climate crisis via awareness campaigns and grassroots movements e.g. Fridays for Future, Sunrise Movement, Extinction Rebellion (*antagonism*)<br><br>Incentivize dirty firms to become clean via:<br>- climate lawsuits, boycotts, and reputational damage (*antagonism*);<br>- institutional reforms, including carbon pricing and mandatory disclosure of risks (*institutionalism*). |

# 4 Overcoming Resistance through Five Strategies

This section reviews the five strategies in detail.

## _Antagonism_

Antagonism springs from grassroots movements by civil society, which aim to awaken public consciousness about climate change, diminish the reputational capital and "social license to operate" of CCCM entities, and pressure governments to act with greater urgency to reduce emissions.

Advocates pursuing this strategy employ tactics which name, shame, sue and boycott the CCCM lobby, thereby increasing the climate consciousness of the citizenry and threatening the business of hydrocarbons. Mass mobilizations, such as those galvanised by Fridays for Future, Extinction Rebellion and the Sunrise Movement fit within the realm of antagonism.

The antagonistic philosophy is well-captured by abolitionist Frederick Douglass' 1857 speech:

> "If there is no struggle there is no progress. Those who profess to favour freedom and yet deprecate agitation are men who want crops without ploughing up the ground; they want rain without thunder and lightning. They want the ocean without the awful roar of its many waters...Power concedes nothing without a demand. It never did and it never will (Douglass 1979, 204). "

In institutional contexts in which there is "political opportunity" (Gamson 1996), that is, a high level of democratization as suggested by citizens having the freedom to assemble, voice demands, exert influence on politicians, and trust the judiciary to remain independent, antagonism may be an effective strategy. One very successful example of antagonism is the Beyond Coal campaign, run by Bloomberg Philanthropies and the Sierra Club, which has retired 60% of US domestic coal-fired power plants (349 out of 530 plants till date)[4] through public awareness and litigation (Sierra Club 2021a; Sierra Club 2021b).

Similarly, condemnatory exposure of alleged wrongdoing can reduce the social license to operate in a business as usual manner. The Exxonknew campaign exposed how the company was aware of the dangers of rising $CO_2$ emissions as early as 1968 but publicly sowed doubt and funded climate denialism, thereby delaying decades of climate action (Oreskes and Conway 2011; Robinson and Robbins 1968). This provided the evidentiary basis for numerous lawsuits filed by states such as New York and California.

Where there is a strong and independent judiciary climate litigation can also be used by citizens against the government. A high-profile case was the Urgenda Foundation _v._ the State of the Netherlands (2019), in which Dutch citizens sued their government over its failure to adopt ambitious climate mitigation measures. The court ruled in favour of citizens arguing that the government was in violation of citizens' constitutional right to secure adequate protection from environmental harm. Such litigation can not only result in direct changes to government policy but also increase how politicians weigh the welfare of climate conscious citizens. There may also be a valid legal case to challenge the issuance of fossil fuel permits when there are low-cost energy alternatives (Rafaty, Srivastav, and Hoops 2020).

---

[4]See: https://coal.sierraclub.org/campaign

*Institutionalism*

Institutionalism involves structural changes to incentivize climate compatible behaviour on a system level. Many institutionalists require "windows of opportunity" to push through their reforms which may arise after elections, mass mobilisations, and exogenous shocks, such as the COVID-19 pandemic, that force the system to do things differently (Farmer et al. 2019). Examples of institutionalist measures that can negatively affect the operations of CCCM corporations include: the establishment of independent climate committees, mandatory disclosure of climate risks, green quantitative easing, conditional bailouts, and negative screens on stock exchanges to ensure listed companies are net-zero compatible (Dafermos, Nikolaidi and Galanis 2018; Hepburn et al. 2020; Farmer et al. 2019). Institutionalism is a strategy best leveraged by those in government, the judiciary, or the technocrats who advise them.

Institutionalism can also involve the establishment of independent oversight committees that shield climate policy from the vagaries of electoral cycles. For example, under the 2008 Climate Change Act, the UK established the Committee on Climate Change (CCC) which was tasked with setting science-based carbon budgets every five years, giving independent advice to the government, and reporting to the Parliament on progress. Independent commissions such as the CCC ensure that there are checks and balances against political short-termism. In many political systems, the creation of arm's length bodies of this sort may be decisive in enhancing the credibility of long-run emissions targets.

*Appeasement*

Appeasement provides compensation to the losers of the transition as a means of quelling their resistance. Leveraging this strategy is typically the prerogative of governments, local authorities, and courts. Common forms of appeasement include worker re-training programmes; pay-offs for workers and asset owners due to early closures of mines; and regional transition funds to support economic diversification in localities that are dependent on climate-forcing assets (e.g. coal, oil, gas, etc.). Appeasement for workers relies on the theory of change that successful strategy uplifts the economic hopes and developmental prospects of low-income communities, fostering a just transition. For example, compensation to miners and their communities was a core element of the climate proposal that US President Joe Biden advanced on the campaign trail when visiting the deindustrialized towns of the Rust Belt.

Appeasement for capital owners, on the other hand, is based on the idea that it may be politically expedient to compensate powerful lobbyists who may otherwise excoriate important reforms - the same way slave-owners were compensated during the abolition of slavery.

Starting in 2015, the Climate Leadership Council (CLC) in the US put forward a national "carbon dividends" proposal that included a provision to establish a legal liability shield, which would statutorily exempt oil and gas companies from all tort liability in court cases seeking restitution for the monetary damages attributed to their historical emissions. This provision was motivated by a theory of change which believed that no comprehensive climate legislation will ever pass through Congress without bringing the oil supermajors to the table. To bring oil supermajors to the table, the policy must not only provide sticks but also carrots (appeasement). This proposal did not prevent the outrage that many environmental groups expressed towards the liability provision. However, there was another segment of environmentalists who preferred to focus on the emissions abatement that could be achieved if "carbon dividends" were adopted. Holding no particularly strong moral conviction about historical liability for emissions, they were willing to endorse the CLC's proposal as a compromise. CLC dropped the proposal in 2019.

_Countervailance_

Countervailance involves supporting green technologies via industrial policy to create a countervailing power to the CCCM lobby. Governments are best placed to leverage the countervailance toolkit through instituting measures such as: R&D tax credits, innovation incubators, subsidies for green innovation, renewable portfolio standards, renewable energy auctions, government procurement for green technologies, and policies that de-risk green investments. The aim of the countervailance toolkit is to increase the uptake of green technologies and bring down their costs so that they can displace carbon-intensive incumbent technologies.

An example of countervailance is Germany's feed-in-tariff for solar energy passed in 2000. One of the authors of the feed-in tariff law argued that history would call it the "Birth Certificate of the Solar Age", since it created assured demand for renewable energy that led to increased production and learning-by-doing (Farmer and Lafond 2016).

Countervailance bypasses head-on engagement with the CCCM lobby and helps dissipate a large portion of the political conflict by enabling market forces to drive rapid deployment (Breetz, Mildenberger and Stokes 2018). As green technologies acquire market share, novel political realignments tend to emerge (Meckling, Sterner and Wagner 2017; Meckling 2019). "Politically active green tech clusters" can become powerful advocates of stronger climate policies, deter policy backsliding, and create further windows of opportunities for institutionalist reform. This feedback dynamic can help advance the energy transition even in the absence of global coordination (Meckling 2019).

An instructive example occurred in Denmark after a centre-right coalition government abandoned several renewable energy commitments in the late 1990s. Vestas, the country's largest wind turbine manufacturer, threatened to leave Denmark and take its suppliers. It formed an ad hoc green lobbying coalition within the Danish Board of Industry. The government quickly realised that it was in its best interest to heed the demands of the green business coalition. They subsequently re-instated various support measures for the wind industry, admitting that they had underestimated the sentiments of big green businesses.

<u>*Co-optation*</u>

Co-optation involves bringing climate policy obstructionists to the side of the climate movement. Co-opters can push for a number of different changes within business organisations such as: commitments to stop funding CCCM lobby groups; linking executive pay to measurable emissions reductions and adopting internal carbon pricing. Co-opters navigate the art and politics of persuasion, and their required skillset is not unlike that of an effective politician.

The theory of change is based on the idea that by convincing a relatively small number of elite individuals, such as the CEOs of large, energy-intensive companies or top government officials, great sums of capital can be reallocated away from climate-forcing assets. Compared to the other strategies, co-optation is available to relatively few members of the climate movement, and perhaps for this reason, its potential is frequently discounted.

Examples of co-opters in the climate movement include Pope Francis who has used his moral authority to summon oil and gas executives to change strategy; family members of executives who are in a unique position to change hearts and minds; and, majority shareholders, high profile advisors, CEOs and elite academics who have a sense of climate consciousness. Co-optation is likely to be a strategy of choice in contexts where ordinary citizens have relatively less bargaining power compared to corporations.

Looking ahead, strategists of co-optation could move beyond attempts to persuade hydrocarbon businesses and start building new alliances with businesses in sectors that have been largely overlooked in climate policy but can play a pivotal role in precipitating change. Google, Amazon, Facebook (Meta) and other technology companies have plans to eliminate or neutralize their carbon footprints. These companies have market-moving power and their actions across supply chains, data centres, and global distribution networks could amplify net-zero efforts in other areas of the economy.

Box 1 gives examples of how the five strategies have been deployed in the climate movement in the US.

## Box 1: US Archetypes of the Five Strategies

Antagonism: Sierra Club *(1892 – present):*

*NGO litigating to close 340+ coal plants across the US*

The Sierra Club, founded in the 19[th] century, uses litigation and grassroots campaigns to decommission coal plants across the US, with 349 plants having closed (amounting to 905 coal-plant production units) and "181 to go" (Sierra Club 2021a; Sierra Club 2021b). The Sierra Club claims to have brought about almost 170MM of clean energy in place of decommissioned coal plants and avoided 2,322 miles of gas pipeline (Sierra Club 2021a).

Institutionalism: Regional Greenhouse Gas Initiative (RGGI) *(2009 – present):*

*A cap-and-trade scheme in Eastern states*

The Regional Greenhouse Gas Initiative Program (RGGI) was the first mandatory, $CO_2$-limiting cap-and-trade programme in the US. Since its inception, the initiative has held 50 auctions, selling 1.11 billion $CO_2$ allowances (worth \$3.78bn in total) to electric power generators in the ten eastern states participating in the program. In 2020, the emissions cap, which drops each year, was 96.2 million tonnes, with an aim of being 86.9 million tonnes in 2030 (Potomac Economics 2010).

Appeasement: The POWER+ Plan *(2016 – present):*

*Compensation to coal communities*

The Obama Administration introduced the POWER+ Plan to invest federal resources in regions that were historically reliant on the coal economy and vulnerable to the energy transition (The White House 2015). The Plan allocated funds to affected workers (\$20m), economic development (\$6m), the Environmental Protection Agency (\$5m), and rural communities (\$97m) (The White House 2016). Since 2015, the Appalachian region in the Northeast (comprised of states like Virginia, West Virginia and Pennsylvania) has received almost \$300m in grants to revive and rebuilt communities (ARC 2021).

Countervailance: California Solar Initiative *(2006 – present):*

*State-wide subsidy for renewable energy*

California launched a \$3.3bn project to subsidise the installation of solar power generation and displace non-renewable fuel sources such as natural gas. To date, the scheme has installed 9,671 MW of solar capacity (about five times the program's initial goal of 1,940 MW) (CPUC 2021; Esfahani et al. 2021).

Co-optation: Climate Action 100+ *(2018 – 2023):*

*An investor coalition using signatories' financial clout to spur climate action*

A coalition of 617 investor signatories with over \$60 trillion in assets under management,[1] the Climate Action 100+ engages with companies to reduce their carbon emissions. Signatories concentrate their lobbying on a "focus list" of 167 companies, who are responsible for an estimated 80% of global industrial emissions. The group's "asks" of companies are to reduce emissions, improve disclosure, and change governance to recognise climate change risk (e.g. in line with TCFD) (Climate Action 100+ 2021).

## 5 Choosing Strategies

We now move to a dynamic perspective and consider how the five strategies build-off each other. Strategy choice depends, in the first instance, on initial conditions related to the macro-structural parameters (democratization, climate consciousness and green business interests) but subsequently, on how the deployment of strategies affects these variables. Therefore, from a dynamic perspective, strategy sequencing is important. To see why, consider the following examples:

*Example 1:*

Consider a setting where the state is heavily captured by business interest groups ($\alpha \approx 0$) and citizens have low climate consciousness ($\beta_1 < 0.5$). This setting could, for example, represent a Middle Eastern petrol producing state. In this context, the strategist will want to focus on increasing the strength of green business interests relative to CCCM interests (i.e., increasing $\phi$) through co-optation or countervailance. Co-optation could be used to convince the ruling elite that global demand for hydrocarbons is likely to diminish and there is a need to diversify towards fast-growing low-carbon industries. Countervailance could play a role in demonstrating the feasibility and disruptive market potential of low-carbon alternatives. Strategies that require political opportunity such as antagonism are unlikely to succeed since $\alpha \approx 0$. If democratic institutional reforms are pursued that increase democratization i.e. $\alpha = 0.25$, then the climate movement's agenda will still face uncertainties since most citizens are against more ambitious climate policy. The pathway in this case would be to first increase green business interests, which may then translate into greater climate consciousness.

*Example 2:*

Let's now consider a case where democratization and green business interests are low but citizens' preferences are tilted strongly in favour of high climate ambition ($\beta_1 > 0.5$). This could be parts of the United States where citizens favour climate action but the political elite is captured by CCCM lobbies. In this case, if a strategist pursues structural democratic reforms (i.e. raising $\alpha$) via institutionalism, then the politician will have stronger incentives to support emissions reductions because the voice of climate conscious citizens suddenly has more weight. In the absence of being able to pursue structural democratic reforms that raise $\alpha$, the climate strategist could continue pursuing co-optation and countervailance to increase green business incentives.

*Example 3:*

Finally, for a strategist in a setting where most citizens favour stronger climate policy and democratization is high (e.g. the Netherlands), there is greater political opportunity through which climate conscious citizens can pursue strong antagonistic tactics such as climate lawsuits. This can directly increase climate policy ambition (e.g. the Urgenda case). The creation of stronger green business interests can also create clusters of green industrial lobbies that can help support institutional reform such as mandatory disclosure of climate risks and countervailance tactics such as subsidies for green technologies.

This simple sketch illustrates how in a dynamic setting, strategies need to be sequenced appropriately since they can build-off each other. Ill-conceived sequencing can lock-in stalemates. There are many potential sequencing options which depend on initial conditions and feedback dynamics.

Strategies may also be deployed jointly to increase efficacy. For example, appeasement on its own, without complementary measures could lead to inefficiently large pay-outs to CCCM capital-owners. This could also create perverse incentives to falsely project continued operations to secure compensation for "early" closures. Germany's coal exit law stipulates that a total of 4.35 billion Euros in compensation will be paid for planned shutdowns by 2030 (Wettengel 2020). However, legal challenges are imminent as the European Commission questions whether "compensating operators for foregone profits reaching very far into the future corresponds to the minimum required" (European Commission 2021). It is likely that antagonism or institutionalism will be needed as complementary strategies to safeguard public interest and put a reasonable upper bound on compensation to capital-owners. Citizens can leverage institutions designed to protect the environment to file antagonistic lawsuits or alternatively, countervailance could be used to create green industrial clusters, which can lobby the government to enact institutional reforms that threaten the CCCM business model.

Our analysis demonstrates that due to positive feedbacks and mutual reinforcement, each strategy likely has a role to play. Some may initially outperform others due to the institutional context, while others may set the stage for more ambitious action subsequently. Tactics that garner the most success are: (i) appropriate to the actors who carry them out; (ii) appropriate to the institutional setting in which they are applied; and (iii) timely.

Previous literature in the field has suggested solutions that fall within one ambit or the other: for example, Meckling et al. (2017) talk about the importance of green industrial policy as a precursor to carbon pricing. By contrast, Zhao and Alexandroff (2019) focus on appeasement as a key strategy, highlighting Germany's compensation efforts as a way to push forward the transition. We combine these perspectives to illustrate how strategy choice and sequencing depend on the initial conditions and the dynamics of three macrostructural parameters: climate consciousness among the citizenry, green industrial incentives and the level of democratization, and how the deployment of strategies in turn also affects these parameters, forming feedback dynamics. Future work could empirically examine how each of these strategies perform in different institutional contexts and explore questions around the sequencing of strategies.

# 6    References

Aguirre, Jessica C. 2021. The Little Hedge Fund Taking Down Big Oil. The New York Times. [online] Available at: <https://www.nytimes.com/2021/06/23/magazine/exxon-mobil-engine-no-1-board.html> [Accessed 18 November 2021].

Aklin, Michaël, and Johannes Urpelainen. 2013. "Political competition, path dependence, and the strategy of sustainable energy transitions." *American Journal of Political Science* 57:3 (September): 643-658.

Appalachian Regional Commission (ARC). 2021. ARC's POWER Initiative. [online] Available at: <https://www.arc.gov/arcs-power-initiative/> [Accessed 18 November 2021].

Ard, Kerry, Nick Garcia, and Paige Kelly. 2017. "Another avenue of action: an examination of climate change countermovement industries' use of PAC donations and their relationship to Congressional voting over time." *Environmental Politics* 26:6 (August): 1107-1131.

Bacon, W. 2013. Sceptical Climate Part 2: Climate Science in Australian Newspapers. *Australian Centre for Independent Journalism*, 1-222.

Bertrand, Marianne, Matilde Bombardini, Raymond Fisman, and Francesco Trebbi. 2020. "Tax-exempt Lobbying: Corporate Philanthropy as a Tool for Political Influence." *American Economic Review* 110:7 (July): 2065-2102.

Blanes i Vidal, Jordi, Mirko Draca, and Christian Fons-Rosen. 2012. "Revolving door lobbyists." *American Economic Review* 102:7 (July): 3731-48.

Bonneuil, Christophe, Pierre-Louis Choquet, and Benjamin Franta. 2021. "Early Warnings and Emerging Accountability: Total's Responses to Global Warming, 1971-2021." *Global Environmental Change* 71: 102386.

Breetz, Hanna, Matto Mildenberger, and Leah Stokes. 2018. "The political logics of clean energy transitions." *Business and Politics* 20:4 (April): 492-522.

Brower, Derek, Ortenca Aliaj. 2021. Engine No 1, the giant-killing hedge fund, has big plans. The Financial Times. 3 June 2021. [online] Available at: <https://www.ft.com/content/ebfdf67d-cbce-40a5-bb29-d361377dea7a> [Accessed 20 November 2021].

Brulle, Robert J. 2014 "Institutionalizing Delay: Foundation Funding and the Creation of US climate Change Counter-Movement Organizations." *Climatic Change* 122:4 (April): 681-694.

Brulle, Robert J. 2018. "The Climate Lobby: A Sectoral Analysis of Lobbying Spending on Climate Change in the USA, 2000 to 2016." *Climatic Change* 149:3 (March): 289-303.

Brulle, Robert J. 2019. "Networks of Opposition: A Structural Analysis of US Climate Change Countermovement Coalitions 1989–2015." *Sociological Inquiry.*

Brulle, Robert J., Jason Carmichael, and J. Craig Jenkins. 2012. "Shifting Public Opinion on Climate Change: An Empirical Assessment of Factors Influencing Concern over Climate Change in the US, 2002–2010." *Climatic Change* 114:2 (February): 169-188.

Burke, J. 2015. Greenpeace bank accounts frozen by Indian government. [online] *The Guardian.* Accessed: 04 November 2020.

California Public Utilities Commission (CPUC). 2021. California Solar Initiative (CSI). [online] Available at: <https://www.cpuc.ca.gov/industries-and-topics/electrical-energy/demand-side-management/california-solar-initiative> [Accessed 17 November 2021].

Cashore, Benjamin, and Steven Bernstein. 2022. "Bringing the Environment Back In: Overcoming the Tragedy of the Diffusion of the Commons Metaphor." *Perspectives on Politics,* 1-24.

Chamon, Marcos, and Ethan Kaplan. 2013. "The Iceberg Theory of Campaign Contributions: Political Threats and Interest Group Behavior." *American Economic Journal: Economic Policy* 5:1 (February): 1-31.

Clark, Cynthia E., and Elise Perrault Crawford. 2012 "Influencing Climate Change Policy: The Effect of Shareholder Pressure and Firm Environmental Performance." *Business & Society* 51:1 (January): 148-175.

Climate Action 100+. 2021. *Investors | Climate Action 100+.* [online] Available at: <https://www.climateaction100.org/whos-involved/investors/> [Accessed 5 December 2021].

Colgan, Jeff, Jessica Green, and Thomas Hale. 2020. "Asset Revaluation and the Existential Politics of Climate Change." *International Organization,* 1-25.

Congressional Research Service (CRS). 2019. *The POWER Initiative: Energy Transition as Economic Development.* Washington, D.C.: Congressional Research Service, pp.1-13.

Dafermos, Yannis, Maria Nikolaidi, and Giorgos Galanis. 2018. "Climate Change, Financial Stability and Monetary Policy." *Ecological Economics* 152 (October): 219-234.

Dal Bó, Ernesto, and Rafael Di Tella. 2003. "Capture by Threat." *Journal of Political Economy* 111:5 (October): 1123-1154.

DellaVigna, Stefano, Ruben Durante, Brian Knight, and Eliana La Ferrara. 2016. "Market-based Lobbying: Evidence from Advertising Spending in Italy." *American Economic Journal: Applied Economics* 8:1 (January): 224-56.

Douglass, Frederick. 1979. *The Frederick Douglass Papers: 1855-63*. Vol. 3. Yale University Press.

Druckman, James N., and Mary C. McGrath. 2019. "The Evidence for Motivated Reasoning in Climate Change Preference Formation." *Nature Climate Change* 9:2 (January): 111-119.

Engine No.1. 2021. Engine No.1 – Homepage. [online] Available at: <https://engine1.com/> [Accessed 20 November 2021].

Esfahani, Asal, Cherie Chan, Christopher Westling, Erica Petrofsky, Joshua Litwin, Narissa Jimenez-Petchumrus, Tory Francisco. 2021. 2021 California Solar Initiative Annual Program Assessment. California Public Utilities Commission (CPUC). [online] Available at: < https://www.cpuc.ca.gov/-/media/cpuc-website/divisions/energy-division/documents/csi-progress-reports/2021-csi-apa.pdf> [Accessed 18 November 2021].

European Commission. 2021. "State aid: Commission opens in-depth investigation into compensation for early closure of lignite-fired power plants in Germany." Press Release. 2 March, Brussels. Retrieved: 14 March 2020 (https://ec.europa.eu/commission/presscorner/detail/en/ip_21_972).

Farmer, J.D. and Lafond, F., 2016. How predictable is technological progress?. *Research Policy, 45*(3), pp.647-665.

Farmer, J.D., Hepburn, C., Ives, M.C., Hale, T., Wetzer, T., Mealy, P., Rafaty, R., Srivastav, S. and Way, R. 2019. Sensitive intervention points in the post-carbon transition. *Science, 364*(6436): 132-134.

Farrell, J., McConnell, K. and Brulle, R., 2019. Evidence-based strategies to combat scientific misinformation. *Nature Climate Change, 9*(3), pp.191-195.

Farrell, Justin. 2016. "Corporate Funding and Ideological Polarization about Climate Change." *Proceedings of the National Academy of Sciences* 113:1 (April): 92-97.

Farrell, J., 2019. The growth of climate change misinformation in US philanthropy: evidence from natural language processing. *Environmental Research Letters, 14*(3), p.034013.

Fink, Larry. 2021. Larry Fink's 2021 Letter to CEOs. BlackRock, Inc. [online] Available at: <https://www.blackrock.com/corporate/investor-relations/larry-fink-ceo-letter> [Accessed 18 November 2021].

Gamson, William A., and David S. Meyer. 1996. "Framing Political Opportunity" edited by Doug McAdam, John D. McCarthy, and Mayer N. Zald." Pp. 275–90 in *Comparative Perspectives on Social Movements: Political Opportunities, Mobilizing Structures, and Cultural Framings*. Cambridge University Press.

Giugni, Marco G., and Florence Passy. 1998. "Contentious Politics in Complex Societies." Pp 81-108. *From Contention to Democracy*. Rowman & Littlefield Publishers, Inc.

Green, Jessica, Jennifer Hadden, Thomas Hale and Paasha Mahdavi. 2021. "Transition, Hedge, or Resist? Understanding Political and Economic Behavior toward Decarbonization in the Oil and Gas Industry." *Review of International Political Economy,* 1-28.

Gross, Samantha. 2019. Mapping Low-Carbon Energy Transitions Around the World: The United States of America. Barcelona: ESADE.

Grossman, Gene M., and Elhanan Helpman. 2001. *Special Interest Politics.* MIT Press.

Gullberg, Anne Therese. 2008. "Lobbying Friends and Foes in Climate Policy: The Case of Business and Environmental Interest Groups in the European Union." *Energy Policy* 36:8 (August): 2964-2972.

Hart, P. Sol, and Erik C. Nisbet. 2012. "Boomerang Effects in Science Communication: How Motivated Reasoning and Identity cues Amplify Opinion Polarization about Climate Mitigation Policies." *Communication Research* 39:6 (December): 701-723.

Hayes, Chris. 2014. "The New Abolitionism". The Nation. Retrieved 16 January 2021. (https://www.thenation.com/article/archive/new-abolitionism/)

Hepburn, Cameron, Brian O'Callaghan, Nick Stern, Joseph Stiglitz and Dimitry Zenghelis. 2020. "Will COVID-19 fiscal recovery packages accelerate or retard progress on climate change?" *Oxford Review of Economic Policy, 36*:1 pp.S359-S381.

Holdo, Markus. 2019. "Cooptation and Non-Cooptation: Elite Strategies in Response to Social Protest." *Social Movement Studies* 18:4 (July): 444-462.

Holyoke, Thomas. 2009. "Interest group competition and coalition formation." *American Journal of Political Science* 53.2 (April): 360-375.

Keohane, R.O., 2015. "The Global Politics of Climate Change: Challenge for Political Science. *PS: Political Science & Politics, 48*(1): 19-26.

Kim, Sung Eun, and Johannes Urpelainen. 2017. "The Polarization of American Environmental Policy: A Regression Discontinuity Analysis of Senate and House Votes, 1971–2013." *Review of Policy Research* 34:4 (July): 456-484.

Leonard, Christopher. 2020. *Kochland: The Secret History of Koch Industries and Corporate Power in America.* Simon & Schuster.

Markard, J. and Rosenbloom, D. 2020. Political conflict and climate policy: the European emissions trading system as a Trojan Horse for the low-carbon transition? *Climate Policy*: 1-20.

McKie, Ruth E. 2019. "Climate Change Counter Movement Neutralization Techniques: A Typology to Examine the Climate Change Counter Movement." *Sociological Inquiry* 89:2 (May): 288-316.

Meckling, Jonas, Thomas Sterner, and Gernot Wagner. 2017. "Policy Sequencing Toward Decarbonization." *Nature Energy* 2:12 (December): 918-922.

Meckling, Jonas. 2019. "Governing Renewables: Policy Feedback in a Global Energy Transition." *Environment and Planning C: Politics and Space* 37:2 (March): 317-338.

Meckling, Jonas and Jonas Nahm. 2022. "Strategic State Capacity How States Counter Opposition to Climate Policy." *Comparative Political Studies.* Vol. 55.

Mercure, J.F., Pollitt, H., Viñuales, J.E., Edwards, N.R., Holden, P.B., Chewpreecha, U., Salas, P., Sognnaes, I., Lam, A. and Knobloch, F. 2018. Macroeconomic impact of stranded fossil fuel assets. *Nature Climate Change, 8*(7): 588-593.

Mildenberger, M. (2020). *Carbon Captured: How Business and Labor Control Climate Politics.* MIT Press

Mildenberger, Matto, and Dustin Tingley. 2019. "Beliefs about Climate Beliefs: The Importance of Second-order Opinions for Climate Politics." *British Journal of Political Science* 49:4 (October): 1279-1307.

Nandi, J. 2020. 3 Green Youth Movements Allege Digital Censorship. [online] *Hindustan Times.* Accessed: 04 November 2020.

Nelder, Chris, host. 2021. "Transition in China". Energy Transition Show (podcast). 06 January 2021. Accessed: 16 January 2021. https://xenetwork.org/ets/

Oreskes, Naomi, and Erik M. Conway. 2011. *Merchants of Doubt: How a Handful of Scientists Obscured the Truth on Issues from Tobacco Smoke to Global Warming.* Bloomsbury Publishing USA.

Potomac Economics. 2010. Annual report on the market for RGGI $CO_2$ allowances: 2009. [online] Available at: <https://www.rggi.org/sites/default/files/Uploads/Market-Monitor/Annual-Reports/MM_2009_Annual_Report.pdf> [Accessed 17 November 2021].

Rafaty, Ryan, Sugandha Srivastav and Bjorn Hoops. 2020. Revoking coal mining permits: an economic and legal analysis. *Climate Policy*: 1-17.

Robinson, E., and R. C. Robbins. 1968. "Sources, Abundance, and Fate of Gaseous Atmospheric Pollutants. Final Report and Supplement." *Stanford Research Institute.*

Rochedo, Pedro RR, Britaldo Soares-Filho, Roberto Schaeffer, Eduardo Viola, Alexandre Szklo, André FP Lucena, Alexandre Koberle, Juliana Leroy Davis, Raoni Rajão, and Regis Rathmann. 2018. "The Threat of Political Bargaining to Climate Mitigation in Brazil." *Nature Climate Change* 8:8 (August): 695-698.

Sierra Club vs. Morton. 1972. U.S. 727 405 (Supreme Court).

Sierra Club. 2021. "*We're Moving Beyond Coal and Gas / Beyond Coal.*" [online] Available at: <https://coal.sierraclub.org/campaign> [Accessed 17 November 2021].

Sierra Club. 2021. *Case Updates.* [online] Available at: <https://www.sierraclub.org/environmental-law/case-updates> [Accessed 17 November 2021].

Skovgaard, Jakob, and Harro van Asselt. 2018. *The Politics of Fossil Fuel Subsidies and their Reform.* Cambridge University Press.

Smelser, Niel. J. 1963. *Theory of Collective Behavior.* The Free Press of Glencoe.

Stokes, Leah Cardamore. 2020. *Short Circuiting Policy: Interest Groups and the Battle over Clean Energy and Climate Policy in the American States.* Oxford University Press, USA.

The White House. 2015. FACT SHEET: The Partnerships for Opportunity and Workforce and Economic Revitalization (POWER) Initiative. Washington, D.C.: The White House, Office of the Press Secretary.

The White House. 2016. Investing in Coal Communities, Workers, and Technology: The POWER+ Plan. [online] Available at: <https://obamawhitehouse.archives.gov/sites/default/files/omb/budget/fy2016/assets/fact_sheets/investing-in-coal-communities-workers-and-technology-the-power-plan.pdf> [Accessed 18 November 2021].

Tong, D., Zhang, Q., Zheng, Y., Caldeira, K., Shearer, C., Hong, C., Qin, Y. and Davis, S.J., 2019. Committed emissions from existing energy infrastructure jeopardize 1.5 C climate target. *Nature, 572*(7769), pp.373-377.

Vesa, Juho, Antti Gronow, and Tuomas Ylä-Anttila. 2020. "The Quiet Opposition: How the Pro-economy Lobby Influences Climate Policy." *Global Environmental Change* 63 (July): 102117.

Wettengel, Julian. 2020. "Spelling out the Coal Exit–Germany's Phase-out Plan." *Clean Energy Wire.* Retrieved: 14 March 2021 (https://www.cleanenergywire.org/news/german-government-and-coal-power-companies-sign-lignite-phase-out-agreement).

Wonneberger, Anke, and Rens Vliegenthart. 2021. "Agenda-Setting Effects of Climate Change Litigation: Interrelations Across Issue Levels, Media, and Politics in the Case of Urgenda Against the Dutch Government." *Environmental Communication* (March): 1-16.

Zhao, Stephen, and Alan Alexandroff. 2019. "Current and Future Struggles to Eliminate Coal." *Energy Policy, 129*, 511-520.